\documentclass[english]{article}
\usepackage[T1]{fontenc}
\usepackage[latin9]{inputenc}
\usepackage{graphicx}
\usepackage[letterpaper]{geometry}
\usepackage{babel}
\usepackage{color}
\usepackage{dsfont}
\usepackage{times}
\usepackage{url}
\usepackage{bbm, amsmath, mathrsfs, float, mathtools, cases}
\usepackage{amssymb}
\usepackage{dcolumn}
\usepackage{multirow}
\usepackage{etoolbox}

\newcommand{\1}{\mathbbm 1}
\newcommand{\sign}{\mathrm{sign}}

\newcommand{\bt}{\mathbf t}

\newcommand{\dd}{\partial}

\newcommand{\qed}{\nobreak \ifvmode \relax \else
      \ifdim\lastskip<1.5em \hskip-\lastskip
      \hskip1.5em plus0em minus0.5em \fi \nobreak
      \vrule height0.75em width0.5em depth0.25em\fi}

\newcommand{\sm} {\setminus}

\begin{document}

\title{A cavity approach to optimization and inverse dynamical problems}

\author{Alejandro Lage-Castellanos$^{1}$, Andrey Y. Lokhov$^{2}$, and Riccardo Zecchina$^{3,4,5}$\\ 
\\
\normalsize $^1$ Department of Theoretical Physics, Physics Faculty, University of Havana, La Habana, CP 10400, Cuba, \\ 
\normalsize $^2$ Universit\'e Paris-Sud/CNRS, LPTMS, UMR8626, B\^at. 100, 91405 Orsay, France, \\
\normalsize $^3$ Department of Applied Science and Technology, Politecnico di Torino, \\
\normalsize Corso Duca degli Abruzzi 24, 10129 Torino, Italy, \\
\normalsize $^4$ Collegio Carlo Alberto, Via Real Collegio 30, 10024 Moncalieri, Italy, \\
\normalsize $^5$ Human Genetics Foundation, Via Nizza 52, 10126 Torino, Italy. \\
}

\date{}

\maketitle

\emph{These are notes taken by A. Lokhov and A. Lage-Castellanos from the review lecture of R. Zecchina given at the autumn school  "Statistical Physics, Optimization, Inference, and Message-Passing Algorithms", that took place in Les Houches, France from Monday September 30th, 2013, till Friday October 11th, 2013.  The lecture covers the work done by A. Braunstein, A. Ramezanpour, F. Altarelli, L. Dall'Asta,  I. Biazzo, A. Lage-Castellanos and R. Zecchina. This material is not intended for publication.\\}

\begin{abstract}
In these two lectures we shall discuss how the cavity approach can be used efficiently to study optimization problems with global (topological) constraints and  how the same techniques can be generalized to study inverse  problems in irreversible dynamical processes. These two classes of problems are formally very similar: they both require  an efficient procedure to trace over all trajectories of either auxiliary variables which enforce global constraints, or directly dynamical variables defining the inverse dynamical problems. We will mention three basic examples, namely the Minimum Steiner  Tree problem, the inverse threshold linear dynamical problem, and the patient-zero problem in epidemic cascades. All these  examples are root problems in optimization and inference over networks. They appear in many modern applications and in a variety of different contexts.  Credit for these results should be shared with A. Braunstein, A. Ramezanpour, F. Altarelli, L. Dall'Asta,  I. Biazzo and A. Lage-Castellanos.
\end{abstract}

\tableofcontents

\section{Introduction}

In 1986 M\'ezard and Parisi treated the traveling salesman problem (TSP) with the tools of statistical mechanics.
The solution given by M\'ezard and Parisi was based on the cavity method (actually the repica method) and predicts the correct expected cost in the large $N$ limit. 
However, a detailed analysis of the equations shows that while the path starts and ends in the same node, it is not constrained to be singly connected. It can be shown that the procedure actually solves the 2-factor problem \cite{frieze02}, being asymptotically equivalent to the TSP \cite{wastlund06}. The neglected global constraint causes a subdominant correction in the cost. On the other hand, any algorithmic treatment of individual instances of the problem has to distinguish between 2-factors and Hamiltonian cycles, and to do so global constraints have to be taken in consideration.

Global constraints are ubiquitous in optimization problems, and, as the previous example showed, they present a major threat to the local iterative methods,
such as the cavity method (also known under the names of belief propagation and sum product equations).
The first of these two lectures shows how to overcome this threat in a set of optimization problems, by translating global constraints into local ones. The passage from global to local constraints is achieved by creating auxiliary variables that undergo an irreversible dynamic in the graph. The method, therefore, can be readily used to study not only global constrained problems, but also irreversible dynamics in graphs. It is presented in the lecture two.

This lecture presents a consistent set of applications of BP to two problems, with growing complexity, that can be found originally in \cite{BBZ2008a,BBZ2008b,biazzo,opt_spread,largedeviations}, together with applications in \cite{pnas:2011,clust_shall,cluster_monkey_cortex,epidemic_bp}. 
We refer to the quoted  papers for a more detailed description of the computations.

\section{Statistical mechanics of optimization with global constraints and Steiner tree problems}

In optimization theory many emblematic problems are stated as the optimization of paths in a weighted graph, as the already mentioned traveling salesman problem. Another example is the minimum spanning tree problem, where we are interested in finding the optimal connected subgraph of a given graph with positive weights in the edges. In what follows we focus in a more general version of the spanning tree, known to be NP-hard, the so-called Prize Collecting Steiner Tree problem  on graphs (PCST) \cite{Johnson2000,Lucena2004}. We use this example as case study to discuss the general issue of  global constraints.

The PCST problem can be stated in general terms as the problem of finding a
connected subgraph of  minimum cost.  Given a graph $G=(V,E)$ with positive  weights $\{c_e:e\in E\}$ on edges
and $\{b_i:i\in V\}$ on vertices, one wants to find a  connected
sub-graph $G'=(V',E')$ that minimizes $\mathcal H(V',E')=\sum_{e\in
E'}c_{e}-\lambda\sum_{i\in V'}b_{i}$, i.e. to compute the minimum:
\begin{equation}
\min_{\begin{array}{c}
E'\subseteq E,V'\subseteq V\\
(V',E')\,\,\mbox{connected}\end{array}}\sum_{e\in E'}c_{e}-\lambda\sum_{i\in
V'}b_{i}.
\label{eq:H}
\end{equation}
the parameter $\lambda$ regulates the tradeoff between the edge costs and vertices prizes, and its value  determines the size of the subgraph $G'$.  It is straightforward  to notice that an optimal  sub-graph must be a tree in that links closing cycles can be removed, lowering the cost. 

This problem is known to be NP-hard and it is encountered in many applications which require its solution as an intermediate step. Examples  range from biology (e.g. finding protein associations in cell signaling
\cite{fraenkel-pcst,pnas:2011}) to technological  network optimization \cite{Hackner:2004}. 

To solve the PCST problem we will adopt  a very efficient heuristics based on belief propagation
developed in \cite{BBZ2008a} that is known to be exact in some limit cases
\cite{BBZ2008a,BBZ2008b}. As in the TSP case, the problem deals with two types of constraints, 
the first one is the minimization of the cost function, and the second one is that the 
resulting subgraph has to be a connected one.

For technical reasons that will be clarified in the next section,  we will deal with a variant of the PCST in which the  optimal tree is required to have a given maximum depth $D$ with respect to a  selected ``root'' node $r$.  This is similar to the so called {\it arborescence} representation  often used in Linear Programming.
Clearly,  a general PCST can be reduced to $D$-bounded rooted PCST by setting $D=|V|$.

\subsection{From global to local constraints}

The cavity formalism requires that  the cost function and the rigid  constraints which appear in the problem  are  written  in terms of local terms.  In the  PCST case the rigid global constraint is the connectivity of the solution.
There are many ways of checking connectivity of a sub-graph. However in order to avoid a check which involves all variables at once one needs to resort an exploration process like a breadth-first-search algorithm. This is an irreversible process in which one visits a node at a time and gain access to its neighbours. 
By mapping the evolution of the exploration process onto a set of distance-to-root variables it is possible to give a local static representation of the connectivity constraint.

We consider a graph $G=\left(V,E\right)$ and a selected \emph{root} node $r\in V$.  Each vertex $i\in V$ is associated to two of variables $\left(p_{i},d_{i}\right)$:
 $p_{i}\in \partial i\cup\left\{*\right\}$, $\partial i=\{j:(ij)\in E\}$ identifies the set of neighbors of $i$ in
$G$ (the parent of $i$) and $d_{i}\in\left\{ 1,\dots,D\right\} $ is the distance of node $i$
to the root node (i.e. the \emph{depth} of $i$).  The special value $p_i=*$  describes the case in  
which node $i$ does not belong to the optimal tree, $i\notin V'$.

To enforce the connectivity condition  and the  tree topology on the subgraph,
variables $p_i$ and $d_i$ are required to satisfy a number of local constraints. 
The depth must decreases along the tree in the direction to the root, i.e.  $p_{i}=j\Rightarrow d_{i}=d_{j}+1$, and 
 nodes that do not participate to the tree ($p_i=*$) must not be parent of some
other node, i.e. $p_i=j\Rightarrow p_j\neq*$. 

For all edges in the graph  we define
$f_{ij}\left(p_{i},d_{i},p_{j},d_{j}\right)=
1-\delta_{p_{i},j}\left(1-\delta_{d_{i},d_{j}+1}(1-\delta_{p_j,*})\right)$
($\delta$ is the Kroenecker delta). 
Then  the subgraph is guaranteed to  be a
tree  if   $g_{ij}=f_{ij} f_{ji}$ equals to one for each edge $\left(ij\right)\in
E$.  We will denote the subgraph $V'$ by ${\mathcal
T}$.

Eventually, by extending the definition of $c_{ij}$ by $c_{i*}=\lambda b_i$, we obtain a fully local definition of the PCST problem:
\begin{equation}
\min{\left\{ \mathcal{H}(\mathbf{p}): (\mathbf{d},\mathbf{p}) \in {\mathcal
T}\right\}},
\end{equation}
where $\mathcal T
= \{ (\mathbf{d},\mathbf{p}):g_{ij}(p_i,d_i,p_j,d_j)=1\,\forall (ij)\in E )$
and 
$\mathcal{H}(\mathbf{p}) \equiv \sum_{i\in V}c_{ip_i}$.

\subsection{Message-passing cavity equations}

The  starting point for the derivation of the cavity equations is the Boltzmann-Gibbs distribution in the extended set of variables:
$P(\mathbf{d},\mathbf{p})= \exp (-\beta \mathcal{H}(\mathbf{p}))/Z$. In the limit $\beta \to \infty$ this probability
concentrates on the configurations which minimize $\mathcal{H}$. 
The BP equations are derived by assuming that the  cavity marginals are uncorrelated and as such satisfy the following  closed set of
equations (see e.g.~\cite{mezard-montanari} for a general discussion):
\begin{eqnarray}
{P}_{j  i}\left(d_{j},p_{j}\right) & \propto & e^{-\beta
c_{jp_{j}}}\prod_{k\in\partial j\setminus i}Q_{k 
j}\left(d_{j},p_{j}\right)\label{eq:phat}\\
Q_{k  j}\left(d_{j},p_{j}\right) & \propto & \sum_{d_{k}}\sum_{p_{k}}P_{k 
j}\left(d_{k},p_{k}\right)g_{jk}\left(d_{k},p_{k},d_{j},p_{j}\right).
\label{eq:bp}
\end{eqnarray}

\begin{figure}
 \begin{center}
\includegraphics[width=0.6\textwidth]{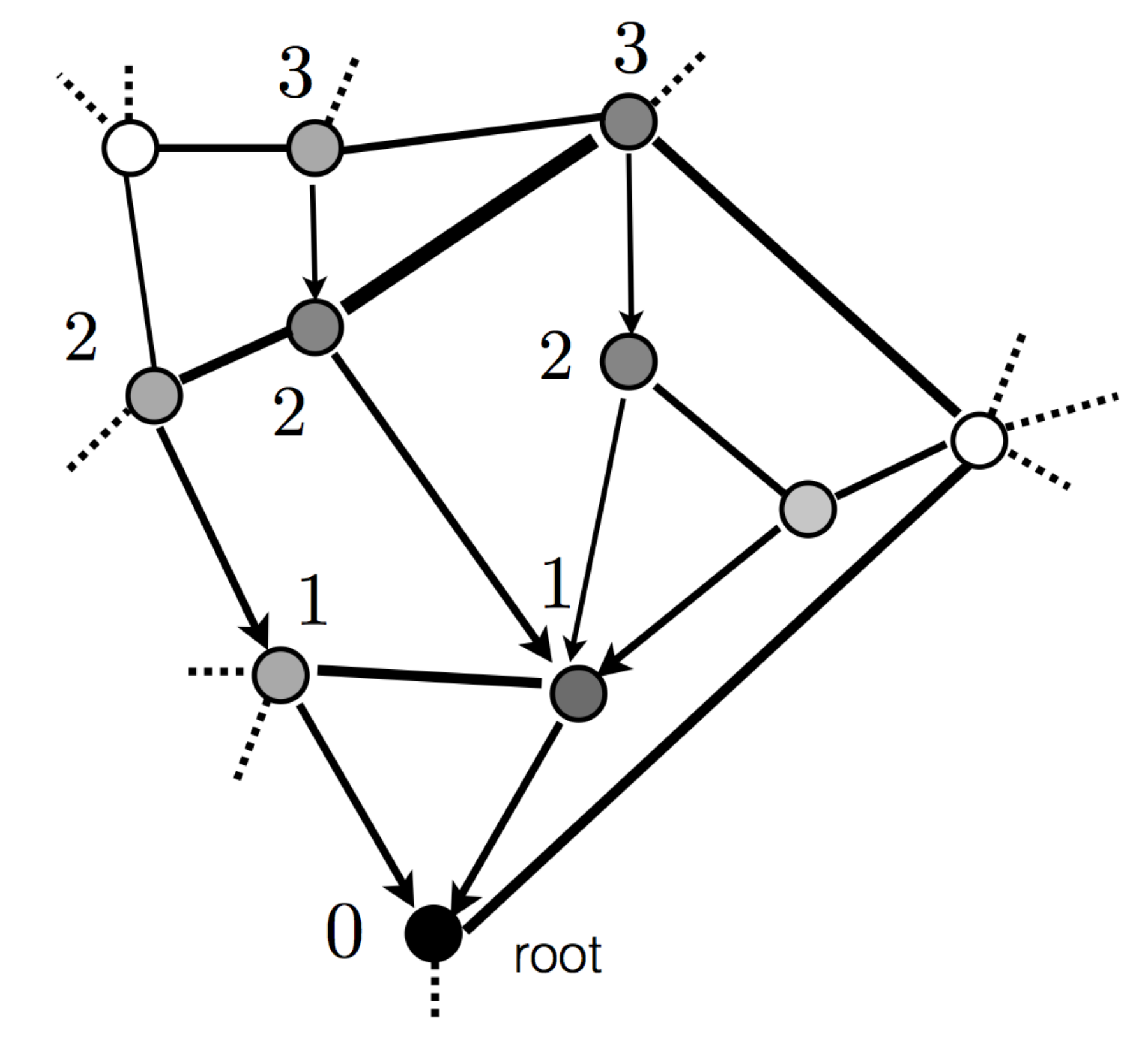}
  \caption{Portion of a graph with a representation of the Prize Collecting Steiner Tree problem. The prize values are proportional to the darkness of the nodes and the edge weights to the thickness of the lines.
Number labels are the distances from the root node.
Distances and pointers are used to define the connectivity constraints.
  }
  \label{fig_stener}
 \end{center}
 \end{figure}

This self-consistent system is iterated until numerical convergence is reached and the cavity marginals are then used to compute the full-marginals under the BP assumtion
\begin{eqnarray}
{P}_{j}\left(d_{j},p_{j}\right) & \propto & e^{-\beta c_{jp_{j}}}\prod_{k\in\partial j}Q_{kj}\left(d_{j},p_{j}\right)\label{eq:bp-marginas}.
\end{eqnarray}

In order to study the  optimization problem we need to take the limit  $\beta\to\infty$, transforming the  BP into the so called Max-Sum (MS) equations.  This is done by  introducing the  cavity field
\[\psi_j(d_j,p_j) = \lim_{\beta\to\infty } \beta^{-1}\log P_j(d_j , p_j)\]
which can be interpreted as the relative negative energy loss of choosing a
given configuration for variables $p_j,d_j$ instead of their optimal choice, 
$$\psi_{j}\left(d_{j},p_{j}\right) = \min{\left\{ \mathcal{H}(\mathbf{p}'):
(\mathbf{d}',\mathbf{p}') \in {\mathcal T}\right\}}-\min{\left\{
\mathcal{H}(\mathbf{p}'): (\mathbf{d}',\mathbf{p}') \in {\mathcal T},
d_j=d'_j,p_j=p'_j\right\}}.$$
In absence of degeneracy, at convergence the optimal values of the fields coincide with the optimal energy in the cavity approximation.

The main problem in using this  iterative scheme as an algorithm resides in the fact that convergence is not guaranteed,
a fact which is observed quite often in practical applications. 
An heuristic way of solving this consists in perturbing the equation by  a technique called reinforcement which amounts at adding a  feedback term in the fixed point Max-Sum equations proportional to the total cavity distribution. This point will be discussed more in detail in section 3.

While the general definition of the PCST problem   is unrooted, the MS equations need a predefined root. 
A trivial way of reducing the unrooted problem to a rooted problem is to solve $N=|V|$ different problems with all possible
different rooting, and choose the one of minimum cost.
A more efficient solution for choosing the "best" root consists in adding a fictitious  new node $r$ to the graph,
connected to every other node with identical edge cost $\mu$. If $\mu$ is
chosen to be sufficiently large then the optimal  energy solution is the  a trivial tree consisting in
just the node $r$. However  the MS  the marginal 
field $\psi_j$ give the relative energy shift of selecting a given parent together with the optimal configuration of the remaining variables. For $\mu$ large enough this energy describes a  tree in which only $j$ 
  is connected to $r$ (as each of these connections costs $\mu$). The smallest energy shift can be used to identify the  optimal root.
In practice one may use this procedure to identify the the best root and successively 
run the rooted MS equations  on the original graph using this choice \cite{biazzo}.

Extensive numerical experiment on renowned benchmark problems as well as a direct comparison with algorithms that define the state of the art in the field (a branch and cut algorithm named DHEA \cite{ivanaConf}, and  a modified version of the Goemans and Williamson algorithm (MGW) \cite{Johnson2000}) show that the performance of the cavity approach  indeed competitive (typically better) \cite{biazzo}.

\subsection{Examples of applications: protein pathways and data clustering}

A direct application of the Steiner Tree problem can be found in \cite{pnas:2011}, where we used it to analyse signalling pathways in large scale transcriptomic and protein interaction data. To test the procedure we showed how to find non trivial undetected protein association in the TOR pathway in {\it S. cerevisiae} as an example. 

One indirect application of the message passing algorithm derived for the Steiner tree is that of data clustering. In the case when the depth of the tree is set to $D=2$ or to $D>n$ (being $n$ the number of elements in the graph) the Steiner tree corresponds to two known algorithms for data clustering, namely affinity propagation \cite{aff_prop} and single linkage \cite{sing_link}. In \cite{clust_shall} we showed how the cavity algorithm improves both methods with intermediate values of $D$, and how this can be applied to the clustering of biological data.

In \cite{cluster_monkey_cortex} we used this clustering procedure to study the visual cortex in monkeys, and concluded that shape accounts more than semantics for the structure of visual object representations in a population of monkey inferotemporal neurons.

\section{From global constraints to inverse problems in irreversible dynamical processes}

The method developed for the Steiner tree problems has the key feature of representing global constraint in terms of  a local irreversible dynamic process of which a static representation is utilized.
The same technique can be adapted to study directly optimization problems  defined over the trajectories of such irreversible processes. These problem are typically characterized by basic units which undergo some irreversible transition from one state to some other state, producing a macroscopic cascade process in system. One may find many examples in a variety of fields.  In physics the most famous examples is certainly Bootstrap Percolation.
In what follows we focus on a simple model called Linear Threshold Model which includes Bootstrap Percolation as a limit case.

The same conceptual scheme can be also generlized to stochastic irreversible processes. In spite of the fact that the approach becomes technically more involved  the computational efficiency can be kept under control. We will just mention at the end the representative application to the patient-zero problem in epidemics.

\subsection{Optimization of trajectories in  spread dynamics}

Consider a  deterministic  dynamics in discrete time defined over a graph $G = (V, E)$ and involving discrete state variables $\mathbf x = \{x_i, i \in V\}$.  We shall assume for simplicity that  there are only two states, $x_i =\in \{0,1\}$ (inactive and active states). 
The process is irreversible, meaning that a vertex that has been activated at some time $t$ will remain active for all subsequent times. For each node the update rule is a function of the  states of the its neighbors in $G$. Here we focus on the Linear Threshold Model (LTM)  \cite{JY05,AOY11,KKT03}, the generalization to other dynamical schemes hopefully being straightforward.

In the LTM  the dynamics is defined by the threshold rule
\begin{equation}
 x_i^{t+1} =
 \begin{cases}
 1 & \text{if } x_i^t = 1 \text { or } \sum_{j \in \dd i} w_{ji} x_j^t \geq \theta_i \,, \\
 0 & \text{otherwise} \, ,
 \end{cases}
\end{equation}
where $w_{ij} $ are positive weights associated to directed edges $(i,j) \in E$,  $\theta_i$ are the thresholds associated to $i \in V$ and $\dd i$ denotes the set of neighbors of $i$ in $G$. 
Clearly this model reproduces the  zero-temperature limit of the random-field Ising model \cite{DSS97,OS10} and  the Bootstrap Percolation process \cite{CLR79,DGM06,BP07}. The active nodes at time $t=0$ are typically called {\em seeds}.

Here we are interested in the  inverse problem of the dynamical evolution, namely the  problem of finding the initial conditions that give rise to a desired final state (results from \cite{largedeviations,opt_spread}). We want to learn how to control the trajectories through initial conditions (but it could also be some external signal at intermediate times).
The example we consider is the objective of finding the set of seeds of minimum cardinality which maximizes the number of active nodes  at some later time T. This problem corresponds to the investigation of the large deviation properties of the dynamics.

\subsubsection{The inverse dynamical problem}

The basic advantage of dealing with irreversibile processes consists in the fact that trajectories to not span an exponential space but are completely parametrized by the unique activation times of the vertices $\bt = \{t_1, \dots, t_N\}$, where $t_i \in \mathcal T = \{0, 1, 2, \dots, T, \infty \}$. We  set $t_i = \infty$ if $i$ does not activate within an arbitrarily defined maximum time $T$.
 
Given a set of seeds $S=\{i: t_i=0\}$,  the dynamical rule for LTM  can be written as  $t_i=\phi_i(\{t_j\})$ with
\begin{equation}
\phi_i(\{t_j\}) = \min \left\{ t \in \mathcal T : \textstyle{\sum}_{j\in\dd i} w_{ji}\1[t_j<t]\geq \theta_i \right\}.
\end{equation}
Admissible trajectories  correspond to strings  $\bt$ such that $\Psi_i = \1 \left[ t_i = 0 \right] + \1\left[t_{i} = \phi_i(\{t_j\}) \right]$ equals $1$ for every $i$. In this settings the functions $\Psi_i$ play the same role as  the constraints $f_{i,j}$ that were used in the Steiner tree problem, while the activation times $t_i$ are similar to the depth variables $d_i$. 

Thanks to this static representation of the trajectories we are free to define an energy function $\mathscr E(\bt)$ that gives different probabilistic weights to different trajectories
\begin{equation}\label{path}
P(\bt) = \frac{1}{Z} e^{-\beta \mathscr E(\bt)}  \prod_{i\in V}\Psi_i(t_i,\{t_j\}_{j\in \dd i})
\end{equation}
with $Z = \sum_{\bt} e^{-\beta \mathscr E(\bt)}  \prod_{i}\Psi_i(t_i,\{t_j\}_{j\in \dd i})$. 
The large deviations properties of the dynamical process can be studied evaluating the static partition function for the dynamic trajectories with an opportunely defined energetic term. The value chosen for $T$ will affect the ``speed'' of the cascade, with lower (larger) values of $T$ corresponding to  the optimization of fast (slow)  trajectories.
\begin{figure}
\begin{center}
\includegraphics[width=.7 \columnwidth]{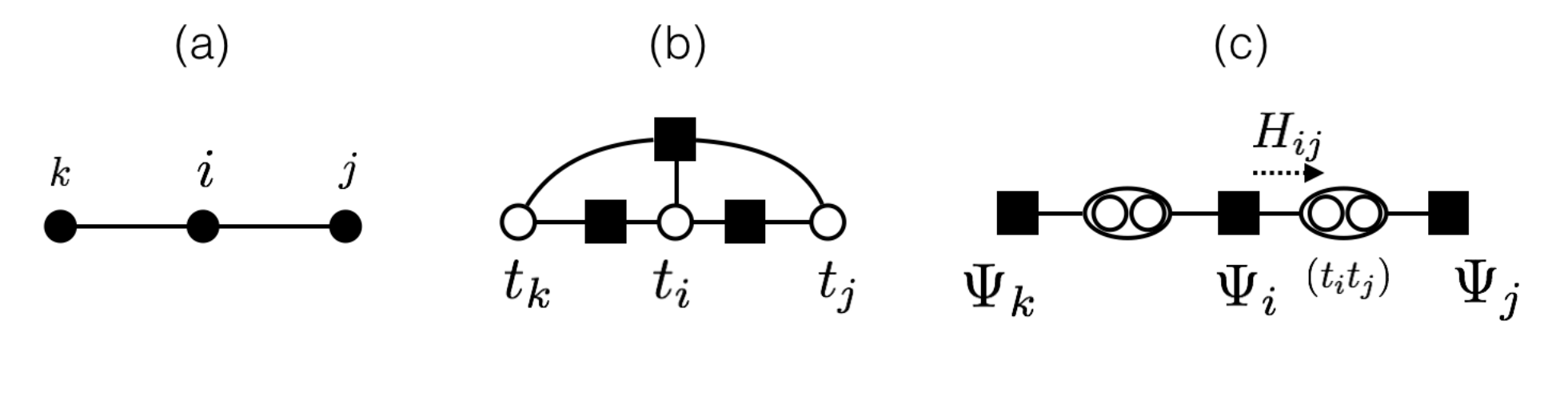}
\caption{ By representing the problem in terms of time pairs ($t_i$ $t_j$), the factor graphs reacquires the original topology of the graph. The factor nodes $\Psi_i, \Psi_j, \Psi_k$   enforce both the dynamical constraints and the consistency of the individual times, namely that the $t_i$ appearing  in ($t_i$ $t_j$)  and in ($t_i$ $t_k$)  coincide.
}
\label{factor}
 \end{center}
\end{figure}
A basic optimization example is the so called Spread Maximization Problem (SMP) where one wants to select the trajectories that are capable of activating the largest number of nodes with the smallest number of seeds. The choice of the energy function can be written as 
\begin{equation}
\mathscr E(\bt) = \sum_i \left\{ c_i \1 \left[t_i = 0\right] - r_i \1 \left[ t_i < \infty \right] \right\},
\end{equation}
where $c_i$ is the cost of selecting vertex $i$ as a seed, and $r_i$ is the revenue generated by the activation of vertex $i$. 
The overall problem is thus defined by the parameters  $\{c_i\}$, $\{r_i\}$, the weights $\{w_{ij}\}$ and the thresholds $\{\theta_i\}$. Trajectories with small energy correspond to an optimal trade-off between seeds cost and  revenue of active nodes at time T.

\subsubsection{Derivation of the Belief-Propagation and Max-Sum equations}

As for the Steiner problem we start by  the finite temperature version ($\beta <\infty$) of the spread  problem for which a Belief-Propagation (BP) algorithm can be used to analyze the large deviations properties of the dynamics  \cite{largedeviations}.

\begin{figure}
\begin{center}
\includegraphics[width=0.6\textwidth]{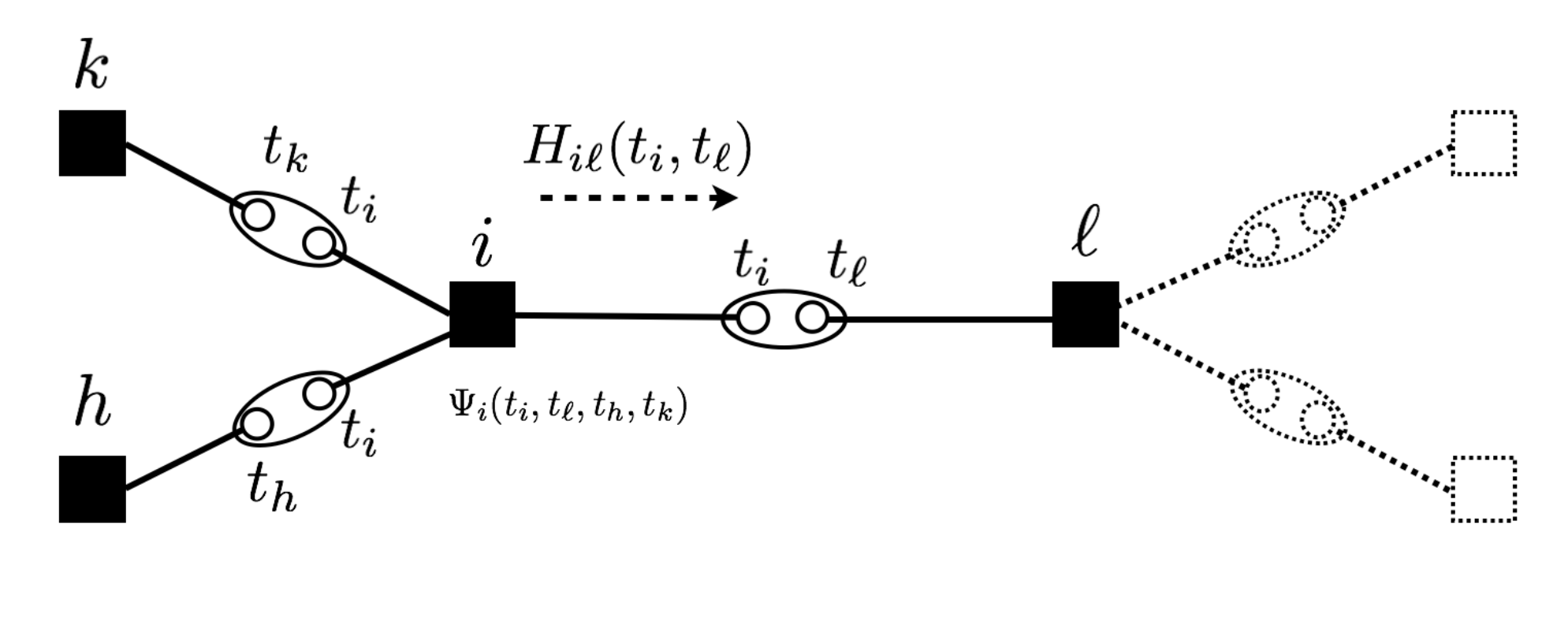}
\caption{Local factor graph representation for spread-like optimization problems}\label{fg}
 \end{center}
\end{figure}

We introduce factor nodes $\{F_i\}$ to represent dynamical constraints $\Psi_i(t_i,\{t_j\}_{j\in \dd i})$ and to each node we associate an energetic contributions $\mathscr{E}_i(t_i)$.  Unfortunately   nearby constraints  $i$ and $j$ share the two variables $t_i$ and $t_j$, a fact that produces  short loops in the factor graph as shown in figure \ref{factor}. 
The local nature of this short loop problem can be solved by writing the equations for the joint cavity marginals of pairs of activation times$(t_i, t_j)$. In this representation variable nodes   are associated to edges $(i, j)\in E$, while the factor nodes are associated to the vertices $i$ of the original graph $G$.  Figure \ref{fg} gives an illustrative example of such "dual" construction.
We call $H_{i\ell}(t_i,t_\ell)$, the marginal distribution for a variable $(t_i,t_\ell)$ in the absence of the factor node $F_{\ell}$. Under the correlation decay assumption, the cavity marginals satisfy the equations \cite{largedeviations}
\begin{equation}\label{bp}
H_{ij}(t_i, t_{j}) \propto e^{-\beta\mathscr{E}_i(t_i)}\sum_{\{t_k\}_{k \in \dd i \sm j}}\Psi_i(t_i,\{t_k\}_{k \in \dd i})\prod_{k \in \dd i \sm j} H_{ki}(t_k , t_i).
\end{equation}
Once the fixed point values of the cavity marginals are found, the ``full'' marginal of $t_i$ can be computed as
$P_i(t_i) \propto \prod_{j \in \dd i} H_{ji}(t_j, t_i)$ and the marginal probability that neighboring nodes $i$ and $j$ activate at times $t_i$ and $t_j$ is given by $P_{ij}(t_i, t_j)\propto H_{ij}(t_i,t_j) H_{ji}(t_j, t_i)$. 
Writing explicitly the constraints and the energetic terms,  the BP equations \eqref{bp} read
\begin{subnumcases}{\label{update1bp} \hspace{-1cm}  H_{i\ell}(t_i, t_\ell) \propto}
  e^{- \beta c_i} \sum_{\{t_j, \, j \in \dd i \sm \ell\}} \prod_{j \in \dd i \sm \ell} H_{ji}(t_j,0)& if $t_i = 0$, \label{update1bp t zero} \\
  \sum_{\substack{
	\{t_j, \, j \in \dd i \sm \ell\} \text{ s.t:} \\
	\sum_{k \in \dd i} w_{ki} \1 [t_k \leq t_i - 1] \geq \theta_i \, , \\
	\sum_{k \in \dd i} w_{ki} \1 [t_k < t_i - 1] < \theta_i
  }}  \prod_{j \in \dd i \sm \ell} H_{ji}(t_j,t_i)   &  if $0 < t_i \leq T$, \label{update1bp t positive} \\
 e^{-\beta r_i } \sum_{\substack{
	\{t_j, \, j \in \dd i \sm \ell\} \text{ s.t:} \\
	\sum_{k \in \dd i} w_{ki} \1 [t_k < T] < \theta_i
  }}  \prod_{j \in \dd i \sm \ell} H_{ji}(t_j,\infty)  & if $t_i = \infty$. \label{update1bp t d}
\end{subnumcases}
The above equations  allow to access the full spectrum of statistics of  the trajectories (e.g. entropies of trajectories or distribution of activation times for given stopping time T and energetic parameters). Very extreme trajectories might require a replica symmetry breaking analysis, a problem which  is beyond the scope of this brief review note.
At the same time, the BP equation do not provide a direct method to explicitly find optimal configurations of seeds (in terms of energy).  To this end, we have to take the zero-temperature limit ($\beta \to \infty$) of the BP equations, and derive the so-called Max-Sum (MS) equations.  As we have learned in the Steiner problem, we introduce the MS messages $h_{i\ell}(t_i, t_\ell)$ defined in terms of the BP messages $H_{i\ell}(t_i, t_\ell)$ as
\begin{align}
h_{i\ell}(t_i, t_\ell) = \lim_{\beta \to \infty} \frac 1 \beta \log H_{i\ell}(t_i, t_\ell).
\end{align}
In  the $\beta \to \infty$ limit of the BP equations \eqref{bp}  one finds
\begin{equation}
h_{i\ell}(t_i, t_\ell) =-\mathscr{E}_i(t_i) + \max_{\substack{\{t_j, j\in \dd i\sm \ell\} \text{ s.t:} \\ \Psi_i(t_i, \{t_j\})=1}}\quad \sum_{j\in \dd i\sm\ell}h_{ji}(t_j,t_i) + C_{i\ell} \label{eq:msfirst}
\end{equation}
that is 
\begin{subnumcases}{\label{update1} \hspace{-0.8cm}  h_{i\ell}(t_i, t_\ell) =}
  \hspace{-0.2cm} \max_{\{t_j, \, j \in \dd i \sm \ell\}} \left[ \sum_{j \in \dd i \sm \ell} h_{ji}(t_j,0) \right] - c_i + C_{i\ell} & \hspace{-0.5cm} if $t_i = 0$, \label{update1 t zero} \\
  \hspace{-0.2cm} \max_{\substack{
	\{t_j, \, j \in \dd i \sm \ell\} \text{ s.t:} \\
	\sum_{k \in \dd i} w_{ki} \1 [t_k \leq t_i - 1] \geq \theta_i \, , \\
	\sum_{k \in \dd i} w_{ki} \1 [t_k < t_i - 1] < \theta_i
  }} \hspace{-0.1cm} \left[ \sum_{j \in \dd i \sm \ell} h_{ji}(t_j,t_i) \right] + C_{i \ell} & \hspace{-0.5cm} if $0 < t_i \leq T$, \label{update1 t positive} \\
 \hspace{-0.2cm} \max_{\substack{
	\{t_j, \, j \in \dd i \sm \ell\} \text{ s.t:} \\
	\sum_{k \in \dd i} w_{ki} \1 [t_k < T] < \theta_i
  }} \hspace{-0.1cm} \left[ \sum_{j \in \dd i \sm \ell} h_{ji}(t_j,\infty) \right] - r_i + C_{i \ell} & \hspace{-0.3cm} if $t_i = \infty$. \label{update1 t d}
\end{subnumcases}
where the additive constant $C_{i\ell}$ is such that $\max_{t_i, t_\ell} h_{i\ell}(t_i, t_\ell) = 0$.
For $t_i = 0$ the maximization is unconstrained and reduces to
\begin{align}
  h_{i\ell}(0,t_\ell) = \sum_{j \in \dd i \sm \ell} \left[ \max_{t_j} h_{ji}(t_j,0) \right] - c_i + C_{ij} \label{update t zero}.
\end{align}
In all other cases  the  updates \eqref{update1 t positive} and \eqref{update1 t d}   require an exponential (in the degree)  number of operations, making a direct implementation unfeasible.
Luckily enough, it is possible to derive  a convolution method and a simplification of both the update equations and the messages  that leads to an efficient updating scheme. This is a crucial and technically involved point which is described in detail in ref.  
\cite{pnas:2011,biazzo}.

\subsubsection{Convergence and reinforcement}
\label{sec:reinf}
In order to force convergence of  MS equations we adopt the so called  the {\em reinforcement} strategy \cite{pnas:2011,BBZ2008a,biazzo} described below . 
The basic idea \cite{BZPRL-2006} is to use the noisy cavity marginals given by MS before convergence to slowly drive the system to a fixed point hopefully without altering the true optimum too much. 

The way this is achieved is by adding an ``external field'' proportional to the total instantaneous local field, with the proportionality constant slowly increasing along with the iterations.
The BP/MS equations  between iterations $\tau$ and $\tau+1$ are modified as follows
\begin{eqnarray}
\hspace{-1.0cm} h^{\tau+1}_{i\ell}(t_i, t_\ell) &=&-\mathscr{E}_i(t_i) + \max_{\substack{\{t_j, j\in \dd i\sm \ell\} \text{ s.t:} \\ \Psi_i(t_i, \{t_j\})=1}}\quad \sum_{j\in \dd i\sm\ell}h^\tau_{ji}(t_j,t_i) + \lambda^\tau p^\tau_{i\ell}(t_i,t_\ell) + C_{i\ell}.\label{eq:msrein}\\
\hspace{-1.0cm} p^{\tau+1}_{i\ell}(t_i, t_\ell) &=& h^{\tau}_{i\ell}(t_i, t_\ell) + h^{\tau}_{\ell i}(t_\ell, t_i) + \lambda^\tau p^\tau_{i\ell}(t_i,t_\ell) + \tilde{C}_{i\ell}
\end{eqnarray}
where $\lambda^\tau = \gamma\tau$ for some $\gamma > 0$ (different increasing functions of $\tau$ are possible). The case $\gamma=0$ corresponds with the unmodified MS equations. We typically observe that the number of iterations scale roughly as $\gamma^{-1}$, while the energy of the solution found increases with $\gamma$. 
This scheme turns BP/MS equations into a concrete solver. While numerical experiments have shown its efficacy, a rigorous understanding of its performance remains to be done.

\subsubsection{Some Numerical Results}\label{sec-result}

The performance of the algorithm have been checked by studying the spread maximization problem on regular random graphs with identical costs $c_i = c$ and revenues $r_i = r$ for all nodes. 
This choice is dictated by the fact that it is known that for this family of graphs optimization problems can be very hard to solve on average \cite{MP} due to the absence of local structural information which can be exploited by simple heuristics.

In order to validate the performance obtained by MS for $\beta \to \infty$, we compared the results with those of three basic strategies, namely  (1) Linear/Integer Programming (L/IP) approaches, (2) Simulated Annealing (SA) and (3) a Greedy Algorithm (GA) (see the original work \cite{opt_spread}). 

The performances obtained by L/IP with the same representation of the problem are extremely poor, becoming practically unfeasible for graphs over a few dozen nodes.  
On the contrary SA and GA algorithms provide results that can be directly used for comparison.
The GA  strategy consists in adding iteratively to the seed set the vertex that achieves the most effective energy improvement. 
Typical solutions are characterized by a final number of seeds that  is substantially larger those obtained using MS. 
Simulated Annealing has better performance than GA but at the price of being  considerably slow. The running time necessary for SA to reach the MS solution scales poorly with the system's size. 
Extensive  numerical have been performed in the case of random graphs of degree $K=5$ and thresholds $\theta=4$ (see ref. \cite{opt_spread} for details).

\subsection{Analytical  solutions for ensembles of random graphs}

The local symmetries  of random regular graphs allow to study the typical solution of the BP equations \eqref{bp} in the single-link approximation \cite{MP}   by the population dynamics method.

In a completely homogeneous setup (i.e.  $w_{ij} = 1$ $\forall (i,j) \in E$, $\theta_i = \theta$, $\forall i \in V$,  $\mu_i =\mu$, $\forall i \in V$ and $\epsilon_i=\epsilon$, $\forall i \in V$), the replica symmetric cavity marginals are expected to be uniform over vertices.
The population dynamics protocol reduces to  a single  self-consistent equation for the  representative BP marginal $H(t,s)$. 
All  links carry the same set of messages and the BP equations can be written  as a relatively simple set of nonlinear equations:
\begin{subequations}\label{BPsl}
\begin{align}\label{BPsl1}
\hspace{-0.5cm} H(0,s) & \propto  e^{-\beta \mu} p_0^{K-1} \\
\label{BPsl2} \hspace{-0.5cm} H(t,s) & \propto \hspace{-0.5cm}  \sum_{\substack{n_- + n_+ + n_0 = K-1 \\ n_- < \theta - \1[s<t-1] \\ \theta-\1[s\leq t-1] \leq n_- + n_0}}  \frac{(K-1)!}{n_{-}! n_{+}! n_{0}!} p_t^{K-1-n_-n_0} m_t^{n_-}  H(t-1,t)^{n_0} \quad \text{for} \quad 0 < t \leq T \\
\label{BPsl3} H(\infty,s) & \propto  e^{-\beta \epsilon} \hspace{-0.5cm} \sum_{n_{-} \leq \theta -1 -\1[s<T]} \binom{K-1}{n_{-}} \left[ H(T,\infty) + H(\infty, \infty) \right]^{K-1-n_{-}} m_{\infty}^{n_{-}}
\end{align}
\end{subequations}
where we introduced the cumulative messages $p_{t} = \sum_{t'\geq t} H(t',t)$ and $m_t = \sum_{t' < t-1} H(t',t)$. 
This  system of equations can be  further simplified by exploiting the fact that $H(t,s) = H(t,\sign(t-s+1))$.
By  varying $\mu,\epsilon,\beta$ and $T$ for any given assignment of $K$ and $\theta$ it is easy to solve numerically the equations \eqref{BPsl}.
On finds that there exist  a large deviation region in the  densities of  seeds for which it is possible to find trajectories that are capable of producing a fully active final state, while typical trajectories which start from randomly chosen seeds get stuck. The entropy of such solution is positive meaning that there exist an exponential number of such optimal trajectories. Complete phase diagrams are reported in ref.  \cite{largedeviations}.

It should be mentioned that the RS solution appears to be correct in a large range of values of the initial densities of seeds, however a precise determination of the minimum optimal density requires a replica symmetry broken computation \cite{guilhem}.

\subsection{Stochastic processes and the patient-zero problem in the SIR model}

As a last remark we mention that in ref. \cite{epidemic_bp} one finds a  further generalization of the cavity approach which deals with the inverse problem for  stochastic irreversible processes.
The archetypal   case which is studied is that of epidemics, in which the state of nodes, at variance with LTM, changes depending of the state of the neighborhood in an stochastic  manner. The forward evolution of the process is not completely defined by the set of active (infected) nodes at time $t=0$, and many evolutions of the process can occur.  A non trivial generalization of the procedure described for the spread dynamics allows to study the so called SIR model of epidemics \cite{epidemic_bp} and to design a novel algorithm for inferring  the most probable origin of an epidemic  cascade given a (possibly noisy)  observation of  the current state at some finite  time $T$. The approximation compares very well to other known methods, being more accurate in most cases.
The formalism can be adapted to study other cases of stochastic irreversible processes.

 
 \bibliographystyle{ieeetr}
 \addcontentsline{toc}{section}{References}
 \bibliography{references}

\begin{thebibliography}{10}

\bibitem{frieze02}
A.~Frieze, ``On random symmetric travelling salesman problems,'' {\em
  Mathematics of Operations Research}, vol.~29, no.~4, pp.~878--890, 2004.

\bibitem{wastlund06}
J.~W{\"a}stlund, ``The limit in the mean field bipartite travelling salesman
  problem,'' {\em preprint}, 2006.

\bibitem{BBZ2008a}
M.~Bayati, C.~Borgs, A.~Braunstein, J.~Chayes, A.~Ramezanpour, and R.~Zecchina,
  ``Statistical mechanics of steiner trees,'' {\em Physical review letters},
  vol.~101, no.~3, p.~037208, 2008.

\bibitem{BBZ2008b}
M.~Bayati, A.~Braunstein, and R.~Zecchina, ``A rigorous analysis of the cavity
  equations for the minimum spanning tree,'' {\em Journal of Mathematical
  Physics}, vol.~49, no.~12, p.~125206, 2008.

\bibitem{biazzo}
I.~Biazzo, A.~Braunstein, and R.~Zecchina, ``Performance of a
  cavity-method-based algorithm for the prize-collecting steiner tree problem
  on graphs,'' {\em Physical Review E}, vol.~86, no.~2, p.~026706, 2012.

\bibitem{opt_spread}
F.~Altarelli, A.~Braunstein, L.~Dall'Asta, and R.~Zecchina, ``Optimizing spread
  dynamics on graphs by message passing,'' {\em Journal of Statistical
  Mechanics: Theory and Experiment}, vol.~2013, no.~09, p.~P09011, 2013.

\bibitem{largedeviations}
F.~Altarelli, A.~Braunstein, L.~Dall'Asta, and R.~Zecchina, ``Large deviations
  of cascade processes on graphs,'' {\em Physical Review E}, vol.~87, no.~6,
  p.~062115, 2013.

\bibitem{pnas:2011}
M.~Bailly-Bechet, C.~Borgs, A.~Braunstein, J.~Chayes, A.~Dagkessamanskaia,
  J.-M. Fran{\c{c}}ois, and R.~Zecchina, ``Finding undetected protein
  associations in cell signaling by belief propagation,'' {\em Proceedings of
  the National Academy of Sciences}, vol.~108, no.~2, pp.~882--887, 2011.

\bibitem{clust_shall}
M.~Bailly-Bechet, S.~Bradde, A.~Braunstein, A.~Flaxman, L.~Foini, and
  R.~Zecchina, ``Clustering with shallow trees,'' {\em Journal of Statistical
  Mechanics: Theory and Experiment}, vol.~2009, no.~12, p.~P12010, 2009.

\bibitem{cluster_monkey_cortex}
C.~Baldassi, A.~Alemi-Neissi, M.~Pagan, J.~J. DiCarlo, R.~Zecchina, and
  D.~Zoccolan, ``Shape similarity, better than semantic membership, accounts
  for the structure of visual object representations in a population of monkey
  inferotemporal neurons,'' {\em PLoS computational biology}, vol.~9, no.~8,
  p.~e1003167, 2013.

\bibitem{epidemic_bp}
F.~Altarelli, A.~Braunstein, L.~Dall'Asta, A.~Lage-Castellanos, and
  R.~Zecchina, ``Bayesian inference of epidemics on networks via belief
  propagation,'' {\em Physical review letters}, vol.~112, no.~11, p.~118701,
  2014.

\bibitem{Johnson2000}
D.~S. Johnson, M.~Minkoff, and S.~Phillips, ``The prize collecting steiner tree
  problem: theory and practice,'' in {\em SODA}, vol.~1, p.~4, Citeseer, 2000.

\bibitem{Lucena2004}
A.~Lucena and M.~G. Resende, ``Strong lower bounds for the prize collecting
  steiner problem in graphs,'' {\em Discrete Applied Mathematics}, vol.~141,
  no.~1, pp.~277--294, 2004.

\bibitem{fraenkel-pcst}
S.-s.~C. Huang and E.~Fraenkel, ``Integrating proteomic, transcriptional, and
  interactome data reveals hidden components of signaling and regulatory
  networks,'' {\em Science signaling}, vol.~2, no.~81, p.~ra40, 2009.

\bibitem{Hackner:2004}
J.~Hackner, {\em Energiewirtschaftlich optimale Ausbauplanung kommunaler
  Fernw{\"a}rmesysteme}.
\newblock na, 2004.

\bibitem{mezard-montanari}
M.~Mezard and A.~Montanari, {\em Information, physics, and computation}.
\newblock Oxford University Press, 2009.

\bibitem{ivanaConf}
I.~Ljubic, R.~Weiskircher, U.~Pferschy, G.~W. Klau, P.~Mutzel, and
  M.~Fischetti, ``Solving the prize-collecting steiner tree problem to
  optimality,'' in {\em ALENEX/ANALCO}, pp.~68--76, 2005.

\bibitem{aff_prop}
M.~Leone, M.~Weigt, {\em et~al.}, ``Clustering by soft-constraint affinity
  propagation: applications to gene-expression data,'' {\em Bioinformatics},
  vol.~23, no.~20, pp.~2708--2715, 2007.

\bibitem{sing_link}
M.~B. Eisen, P.~T. Spellman, P.~O. Brown, and D.~Botstein, ``Cluster analysis
  and display of genome-wide expression patterns,'' {\em Proceedings of the
  National Academy of Sciences}, vol.~95, no.~25, pp.~14863--14868, 1998.

\bibitem{JY05}
M.~Jackson and L.~Yariv, ``Economie publique,'' {\em Numero}, vol.~16,
  pp.~3--16, 2005.

\bibitem{AOY11}
D.~Acemoglu, A.~Ozdaglar, and E.~Yildiz, ``Diffusion of innovations in social
  networks,'' in {\em Decision and Control and European Control Conference
  (CDC-ECC), 2011 50th IEEE Conference on}, pp.~2329--2334, IEEE, 2011.

\bibitem{KKT03}
D.~Kempe, J.~Kleinberg, and {\'E}.~Tardos, ``Maximizing the spread of influence
  through a social network,'' in {\em Proceedings of the ninth ACM SIGKDD
  international conference on Knowledge discovery and data mining},
  pp.~137--146, ACM, 2003.

\bibitem{DSS97}
D.~Dhar, P.~Shukla, and J.~P. Sethna, ``Zero-temperature hysteresis in the
  random-field ising model on a bethe lattice,'' {\em Journal of Physics A:
  Mathematical and General}, vol.~30, no.~15, p.~5259, 1997.

\bibitem{OS10}
H.~Ohta and S.-i. Sasa, ``A universal form of slow dynamics in zero-temperature
  random-field ising model,'' {\em EPL (Europhysics Letters)}, vol.~90, no.~2,
  p.~27008, 2010.

\bibitem{CLR79}
J.~Chalupa, P.~L. Leath, and G.~R. Reich, ``Bootstrap percolation on a bethe
  lattice,'' {\em Journal of Physics C: Solid State Physics}, vol.~12, no.~1,
  p.~L31, 1979.

\bibitem{DGM06}
S.~N. Dorogovtsev, A.~V. Goltsev, and J.~F.~F. Mendes, ``K-core organization of
  complex networks,'' {\em Physical review letters}, vol.~96, no.~4, p.~040601,
  2006.

\bibitem{BP07}
J.~Balogh and B.~G. Pittel, ``Bootstrap percolation on the random regular
  graph,'' {\em Random Structures \& Algorithms}, vol.~30, no.~1-2,
  pp.~257--286, 2007.

\bibitem{BZPRL-2006}
A.~Braunstein and R.~Zecchina, ``Learning by message passing in networks of
  discrete synapses,'' {\em Phys. Rev. Lett.}, vol.~96, p.~030201, 2006.

\bibitem{MP}
M.~M{\'e}zard and G.~Parisi, ``The bethe lattice spin glass revisited,'' {\em
  The European Physical Journal B-Condensed Matter and Complex Systems},
  vol.~20, no.~2, pp.~217--233, 2001.

\bibitem{guilhem}
A.~Guggiola and G.~Semerjian, ``{Minimal contagious sets in random regular
  graphs},'' {\em arXiv:1407.7361}, 2014.

\end{thebibliography}

\end{document}